# Challenges in low losses and large acceptance ion beam transport

F. Osswald,[a,*] E. Traykov,[a] T. Durand,[b,c] M. Heine,[a] J. Michaud,[d] JC. Thomas[e]

[a] *IPHC, CNRS/IN2P3, Université de Strasbourg, 23 rue du Loess, Strasbourg, France*
[b] *SUBATECH, CNRS/IN2P3, IMT Atlantique, Université de Nantes, 4 rue Alfred Kastler, Nantes, France*
[c] *GIP ARRONAX, 1 rue Aronnax, Saint-Herblain, France*
[d] *LP2IB, CNRS/IN2P3, 19 chemin du Solarium, Gradignan, France*
[e] *GANIL, CEA/DRF-CNRS/IN2P3, Boulevard Henri Becquerel, Caen, France*

*Email:* francis.osswald@iphc.cnrs.fr

ABSTRACT: A prototype of ion beam transport module has been developed at the Institut Pluridisciplinaire Hubert Curien (IPHC) and used as a test bed to investigate key issues related to the efficient transport of ion beams. This includes the reduction of the beam losses, the increase of the acceptance, and the definition of the instrumentation necessary to evaluate the performances. An experiment was performed on a full-scale beam line and following a standard beam analysis, steering, and focusing procedure. After a review of the developments carried out for some demanding facilities and for the design of the quadrupoles implemented in the transport module, the paper highlights the challenge of measuring the preservation of transverse phase-space distributions with large acceptance conditions, i.e. with the highest ratio of beam filling to quadrupole aperture. Then, the tolerance to the errors and mitigation of the risks are discussed, in particular by considering the electric stability of the transport module, beam trips, behavior of the tail and the halo, and misalignment errors.

___________________________________
* Corresponding author.

# Contents

## 1. Introduction

Due to increasing environmental and economic constraints, optimization of ion beam transport and equipment design becomes essential. The future should be equipped with planet-friendly facilities, that is, solutions that reduce environmental impact and improve economic competitiveness. The tendency to increase the intensity of the current and the power of the beams obliges us and brings us to new challenges. Installations tend to have larger dimensions with increased areas, volumes, weights and costs. This challenge is achievable by reducing emittance growth and limiting the aperture of the equipment. The emittance growth limitation reduces beam losses, and equipment with reduced aperture limits the costs, but the two objectives are antagonists because a good transmission is generally obtained with a large aperture. Previous works highlighted the importance of the operational radiation protection measurements allowing a realistic estimate of the radioactivity induced by beam losses, see for example [1]. In the case of the LINAC4, the estimation is based on a 0.1 W/10 m beam loss assumption for an irradiation profile over 30 years. The induced radioactivity is produced by direct activation of the accelerated $H^-$ anions and by neutron induced activation occurring during commissioning, ramp-up, and operation. The beam losses depend on the aperture of the equipment and reach $10^{-3}$ of the total average intensity. To complete beam losses estimation, it is customary to consider the beam envelope, the distribution of the charged particles, and also the discrepancies produced by beam fluctuation, trips, misalignment, and mis-setting, i.e. error scenarii. In the case of powerful accelerators and high intensity RIBs, these "unpredictable" events require the consideration of margin and tolerances. For the new JAEA-ADS designed for 30 MW proton beam, the standard 50-75% filling of the quadrupole aperture (at 3 rms for a Gaussian distribution) is pushed to 15 rms to reduce beam losses [2]. This issue will be discussed and illustrated by some experimental results in the following. For this purpose, a new focusing unit based on a quadrupole doublet structure has been constructed at the Institut Pluridisciplinaire Hubert Curien (IPHC). The prototype was developed to study some key issues for the transport of low energy ion beams and can be adapted to other conditions - it is based on electrostatic technology but can be converted into electromagnets. Similar theoretical developments have been made for both technologies to define the aberrations produced by the distributions of electrostatic potential and magnetic scalar potential as detailed in the following. The prototype used during our experiment has a 0.5 m long quadrupole doublet structure. The design led to a new pole shape based on field maps optimization and high-order modes analysis (HOM). The design will reduce optical aberrations produced by nonlinear fields that disturb the outer part of the beams, thereby improving the transport properties thanks to the reduction of the emittance growth. The typical application is the transport of radioactive ion beams (RIBs), see SPIRAL 2 and DERICA projects for example [3-5]. Despite the low current intensity (less than 10 nA) and standard loss level of $10^{-4}$, cumulative contamination can limit access and reduce operability [6]. There are many similar applications requiring enhanced beam transport, for example at SNS for the test facility [7], for a focusing channel designed at RIKEN [8] and used at LANL [9], at IHEP for a multipole halo suppressor [10] and for Accelerator Driven Reactor Systems [11], at FRIB for the ARIS magnetic separator [12], at KIT [13] and FAIR for storage rings with large acceptance [14], and at Daresbury for Laser Wakefield Acceleration [15]. Finally, the application for miniature accelerators is the new trend. The same design principle allowing large acceptance in the transverse trace-space (in position and angle) can also be used to reduce the equipment size, see Fig. 1 with two different bore radii and same beam.

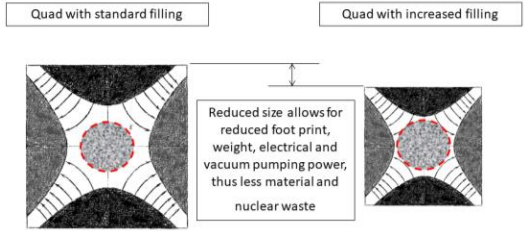

**FIG. 1.** Principle of the quadrupole with a large acceptance and reduction in size. Left: beam centered in a uniform field region, with standard filling (50-60%), reduced harmonic composition, and beam losses for long beam lines. Right: increased filling (80-90%), enlarged acceptance, non-uniform fields, presence of high-order harmonics, reduced emittance growth due to optimized pole shape, low beam losses, and reduced manufacturing and operation costs.

An experimental campaign conducted by IPHC and IN2P3 laboratories was carried out in order to analyze key problems related to the design and operation of low losses and large acceptance beam transport. The results and the difficulties encountered in characterizing the performance of the doublet are reported with emphasis on low current intensity measurement (tail and halo of the beam), 2D transverse emittance figures with non-standard elliptical shape, background noise contribution, control of experimental conditions such as beam stability, power supply regulation, and beam settings. The method to evaluate the beam transport performances with a filling of the quadrupole up to 80% will be described as well as other side effects related to absolute emittance measurements.

## 2. Design of the poles

### 2.1 High-order modes analysis

The prototype is an operational beam transport module with an optical structure based on an electrostatic quadrupole doublet (FODO structure with successive quadrupoles noticed Q1 and Q2). The design and manufacture of the prototype and especially the poles were carried out under a research contract between the CNRS and the Sigmaphi company in France. The design of the poles is based on an adequate ratio between the radius of the pole, their shape, the radius of the bore, and the length. With a quadrupole, the radial field has a linear expansion in the central region between the poles and becomes nonlinear near the pole tips due to their axial finite size. The quadrupole pole shape design prototype is carried out following an analysis of the fields and a decomposition by Fast Fourier Transform (FFT) of the higher-order multipole fields, as described in [16]. The field harmonics do not depend on the radius but on the axial and the azimuthal positions. However, the coefficients of the Fourier series decomposition vary with radial position. The harmonics and their coefficients are extracted from the field map, integrated on the axial position with the OPERA3D code in order to evaluate their importance over the full length of the doublet (optical structure and fringe field) with respect of the main quadrupole component $A_2$. In the case of a quadrupole, the various multipoles $A_n$ are referenced by their azimuthal position, and the index n = 4m-2 defines the harmonics $A_6$, $A_{10}$, and $A_{14}$ allowed by the symmetry can be calculated in addition to that of the quadrupole $A_2$ (m index ≥ 1).

Higher-order field components generate optical aberrations and limit beam transport. The oscillations of the outer particles near the tips of the poles are non-harmonic while those near the center in the uniform field remain paraxial with a constant betatron wavelength. Preliminary developments in the domain were limited by two-dimensional models (or calculations in the median plane), underestimating effects of fringe field, and with Gaussian and first order beam optics approximations. This allowed the shaping of the electrostatic and magnetic poles in 2D mainly with circular and hyperbolic shapes and the production of uniform fields over the greater part of the length [17-26]. For example, the hyperbolic shape with a finite section generates higher-order components at the tips but their contributions are neglected, and the field gradient can be considered small. Moreover, due to the axial extension, more consistent contributions to the field gradient are omitted and limit the relevance of the design, see for example [22]. With the advent of numerical analysis and high-performance computing in the 90's, the design of 3D models, the detailed calculation of high-order field components and optical aberrations have become accessible. More recently, some works questioned the conventional design of electrostatic and magnetic quadrupoles and proposed complementary approaches in order to estimate more accurately the exact 3D geometry for both the technologies, the ends, and the fringe field [27-32]. This involves the reduction of several multipole components related to the azimuthal harmonics generated by the quadrupole symmetry, more specifically the integral of $A_6$ over the axial direction, and the high-order aberrations (third and higher). The different coefficients and the first, third and fifth order forces/aberrations can be easily calculated with different codes as Cosy-infinity, GIOS, OPERA3D, TRACE-3D, etc. with more or less precision. The usual observed main effect is the third order aberration which is in fact a cubed power variable (i.e. the transverse dimension x) supplemented by a dependence with the xy² factor and an inverse dependence with the effective length and with the power squared focal length.

After fabrication of the doublet, control of the resulting field in the aperture and near the surface of the quads is difficult due to the electrostatic influence of the probe and the desired precision. Since then, quality assurance based on a rigorous machining process and a 3D control procedure for parts and assemblies have been implemented. For example, a multiaxial arm was used to control the position of the assembled poles and the holding structure inside their vacuum chamber with a tolerance of ±100 µm (FARO equipment). Nevertheless, not everything designed can be achieved at reasonable costs with a conventional CNC machining center. Prospects for reducing costs (especially personnel costs due to production time) and allowing the manufacturing of complex pole shapes are emerging thanks to recent advances in the field of additive manufacturing, see for example [33-34].

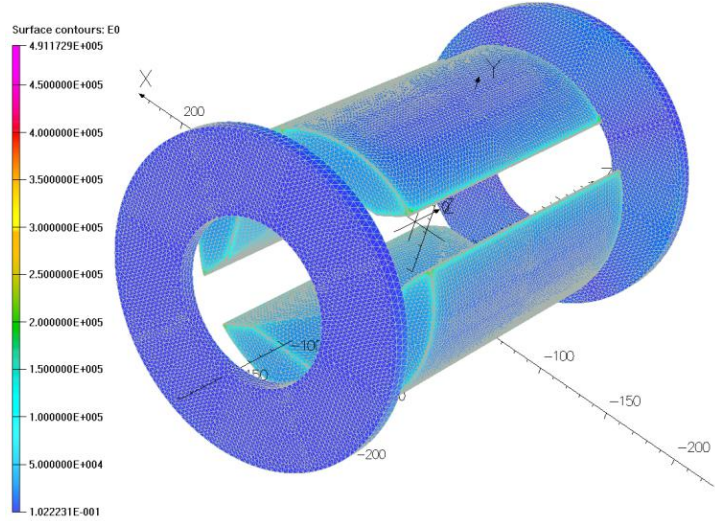

**FIG.2.** Numerical 3D model with field map on the surfaces of the quadrupole and two collimator rings.

**2.2 Design optimization**

The principle of the method consists in minimizing some of the integrated high-order field components at a radial position, typically chosen to 50-75% and here 80% of the aperture because this corresponds to the filling of the beam. Then, the beam optics is obtained after integrating the equations of motion of the particles along the length of the doublet. The design of the poles is carried out with a new iterative method, improved over time, and based on several routines including parametric modeling and optimization of the OPERA3D code. Pole design optimization is performed with specific objective functions and design variables in the Optimizer module of the OPERA3D code, developed at IPHC, and allowing automatic tasks and time savings. Algorithms dedicated to multi-objective optimization problems for the design of particle accelerator magnets are relatively rare and often limited to 2D models (with longitudinal invariance of the quadrupole cross-section), see for example [35]. Our objective functions are defined by the integrated quadrupole component ($A_2$) which must be kept constant and several unwanted HOMs which must be minimized. Unwanted HOMs are defined by their negative impact on particle trajectories. For example, some components have a high-order radial dependence, with a radial field proportional to the radius to the fifth power and higher. It is a determining factor for particles passing close to the tips and for the beam halo which can be used as a signature of the performance. Precise termination criteria and numerical limits of design variables are defined for this purpose and avoid divergence during the iterations. New routines have been developed since the first design based on hyperbolic poles, cut offs, and fine shimming [16]. A new script has been defined recently to build a volume showing a smooth profile from "smoothing" operations based on sweeps, extrusions and cuts following a predetermined number and size of steps and within certain limits (the geometric constraints), see Fig. 3. The construction is controlled by the minimization of a few integrated components with OPERA3D specific to the fourfold symmetry, mainly the dodecapole (12 poles, component $A_6$), the duodecapole (20 poles, component $A_{10}$), and the 28 poles (component $A_{14}$) because they tend to defocus the beam and lead to aberrations. Then, an experimentation plan is defined in order to allow a comprehensive exploration of the different parameters i.e. to optimize the search of the best solution in a minimum of time.

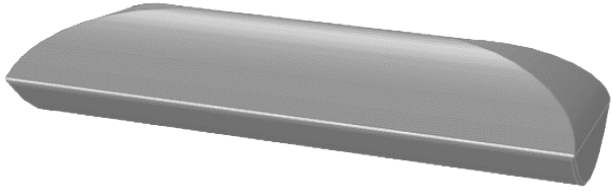

**FIG. 3.** Optimized pole design obtained from the original shape and additional geometric constraints leading to smoothing the profile and minimizing some harmonics. The shape is the result of an experimentation plan which takes 1 hour on a standard laptop (64 bits, 2.8 GHz, 32 Go RAM).

**2.3 Beam optics**

The design of the prototype quadrupoles follows an iterative and interactive process with geometry variation, field harmonics analysis, and beam optics calculation. A staged simulation approach has been applied to three successive models: a single quad with ideal environment in order to evaluate the raw shape of the pole, a quad doublet positioned in the real environment with the mechanical structure and the vacuum chamber to obtain the fields and dynamics of the beam, and finally a virtual beam line with a fixed number of identical quad doublets in order to amplify the defects and observe detailed behavior of the optical structure over a long distance. Beam optics calculations are performed with TraceWin code [36]. To note, the multi particle mode of the code (also called PATRAN mode) allows the evaluation of high-order beam optics and therefore the effects of the HOM components on the aberrations (beam matrix calculations are generally limited to first and second order beam optics). The result of the transverse distributions in 2D phase-space for a 34 m long virtual beam transport line is shown in Fig. 4. The optical structure is composed of six-unit cells of identical doublets. The initial Gaussian distributions have a marginal emittance of 80 $\pi$ mm.mrad and are truncated at ±3 rms in the transverse planes. With the optimized design the rms emittance growth is less than the usual 1% in both transverse planes, the variation of the rms beam radii $x_{rms}$ and $y_{rms}$ is about 0.5% and the transmission is better than 99.5% with a quad filling of 80% [6]. Due to the quadrupole symmetry, there are no second order aberrations and the filamentation of the emittances is due to the 3$^{rd}$ order aberrations i.e. the nonlinear fields applied to the charged particles of the tail and halo. The effect is typical of transport lines but can be observed on periodic and circular structures.

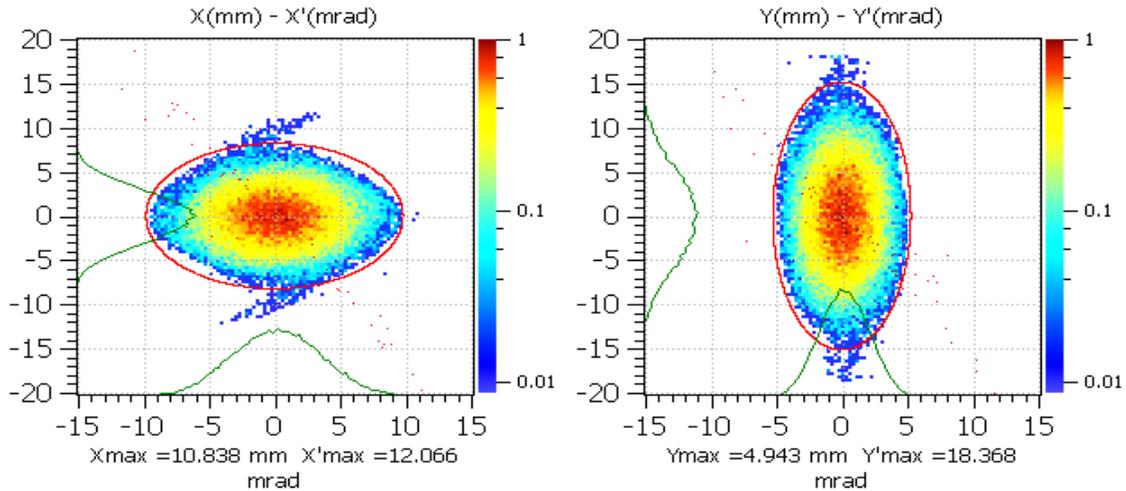

**FIG. 4.** Transverse trace-space distributions simulated in X-X' and Y-Y' planes at the exit of a 34 m long virtual beam transport line. The red ellipses indicate the 3-rms emittances. The charged particles are subjected to nonlinear fields near the pole tips. These particles will be lost in the beam line after a certain distance (the simulation generates a population of $10^5$-$10^6$ particles).

## 3. Experimental investigations

High performance transport requires preservation of the transverse phase-space distributions and large acceptance conditions, i.e. with the highest ratio of beam filling to quadrupole aperture. Nevertheless, under these conditions, new constraints appear, and given the tolerances of the system, unstable behavior can occur. The aim of the experiment is to investigate the challenge of the emittance growth measurement with different perturbations. Differential measurements are carried out with two emittance scanners in order to evaluate the emittance growth for different beam sizes, i.e. for several fillings of the first quadrupole aperture of the doublet. Measuring the transverse phase-space distributions is crucial and requires high resolution instruments in order to detect the tiny signature of the HOM components on the beam tail and halo. The lower the disturbance of the initial emittance ellipse, the better the transport of the beam. Both emittance scanners were calibrated in a preliminary experiment [37]. The first device is a commercial 4D pepperpot system from Pantechnik (EM1), and the second one is an "Allison" type 2D scanner (EM2) developed at IPHC [38]. The stability of the transport module with beam trips and misalignment errors play an important role in the beam losses and received special attention during the experiment.

### 3.1. Experimental setup

The experiment is performed at the ARIBE test facility operated by the CIMAP laboratory and GANIL, Caen. The beamline presents the configuration of a standard low energy ion beam line and delivers an analyzed $^{40}Ar^{8+}$ beam (without contaminants). It offers a variety of beam instrumentation equipment allowing for the different beam settings and the production of large beam transverse dimensions (up to 100 mm in diameter). The test bench with the quadrupole doublet structure was installed at the extremity of the beam line. The facility delivers a 15 keV/q $^{40}Ar^{8+}$ beam with an intensity typically between 100 nA and 1 µA DC. The beamline is equipped with an ECR ion source, several dipoles (D), quadrupoles (Q), and steering magnets. The experimental setup including a vertical slit (FV), the quadrupole doublet (Q1 and Q2), the emittance meters (EM1 and EM2), and their location are shown in Fig. 5. Two strategies of beam matching were considered: the first one is based on a standard point to point steering and focusing from upstream to downstream, and the second one consists of a memory-based setup with additional iterative optimizing i.e. initial conditions complemented by small steps. The simulations carried-out with TraceWin offer some references for comparisons with the measurements and allow a better understanding of the discrepancies and the perturbations.

Due to the importance of the beam alignment on the emittance growth, the F22 – F24 slits were systematically used to redefine and center the beam on axis in both planes before the D4 dipole. Beam centering was checked using the PR23, PR41, and PR42 beam profilers. The F22-F24 slits have been set to very small apertures in the horizontal plane in order to reduce the horizontal emittance, thus limiting the contribution to the measurements. This was necessary due to the use of the vertical slit (FV) and the EM1 and EM2 emittance meters which were installed to measure the vertical beam emittance only. FV slit combined with a defocused beam upstream allows the selection of a uniformly distributed beam sample for the purpose of the quadrupole aperture filling. The settings of Q1 and Q2 quads are fixed in order to allow successive focusing on the two emittance scanners allowing a comparison of the measurements, and the evaluation of the emittance growth from an unperturbed beam (with a lower filling). The settings of Q2 are standard, with fillings less than 80%, and allow an adequate focusing on the emittance-meters located downstream.

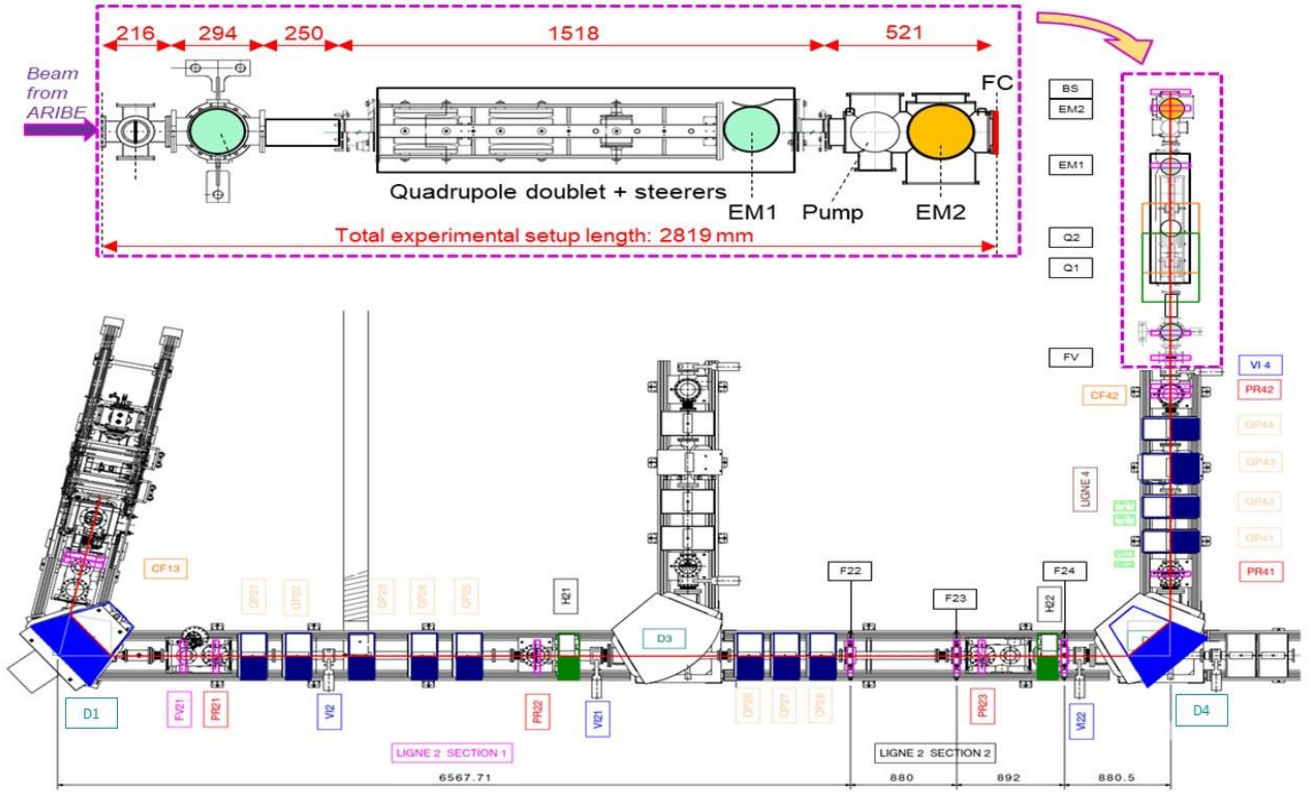

**FIG.5.** Experimental setup on the ARIBE-D4 beamline. The ECR ion source is positioned before the first dipole magnet (D1) on the left side of the beamline, followed by two dipole magnets (D3 and D4) and other equipment as slits ($F_i$), profilers ($PR_i$), and Faraday cups ($CF_i$). The end of the line is identified by a vertical slit (FV) at the interface with the test bench (prototype with a quadrupole doublet structure). The FV slit enables to control the filling with a uniform distribution of the first quad of the doublet structure, and therefore the large acceptance conditions of the experiment. The beam stop at the end of the beam line is replaced by a Faraday cup (FC).

## 3.2 Control of the quadrupole filling and transverse distribution

Of particular importance is the control of the beam filling and the uniformity of the transverse distribution. The first criterion is necessary to guarantee large acceptance measurements, the second allows higher current intensities and therefore better precision for measurements in the tail and halo. For the experiment, quadrupole filling must be evaluated with precision during each emittance growth measurement. It is defined as the fraction of the aperture between the opposite poles at the entrance of the first quadrupole (±50 mm). It is fixed by the vertical slit aperture (FV), and is evaluated by the geometry and the vertical beam waist at the position of PR42. The evaluation is complemented with TraceWin simulations to define the several beam optics settings, see Fig. 6. The uniform transverse distribution in the vertical plane is obtained by defocusing of the beam and selection of a central sample with the FV slit.

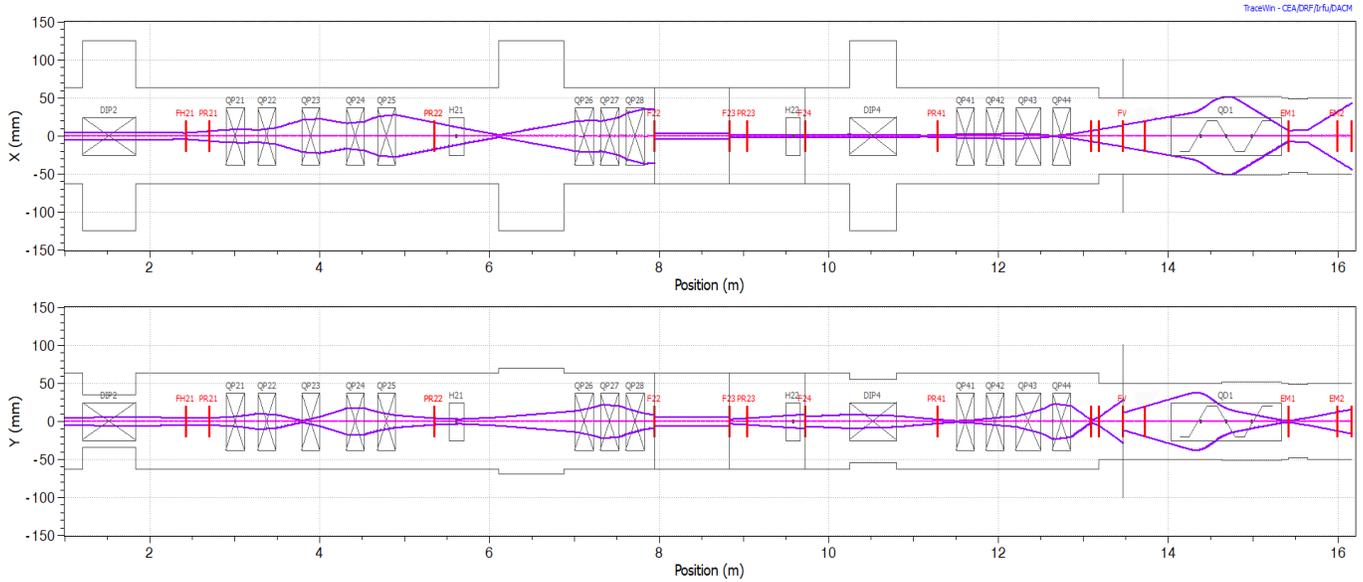

**FIG. 6.** Beam optics with rms envelopes in both transverse planes. Beam matching is mainly obtained with the quadrupoles (QP21-QP44) and the emittance matching section composed of three slits (F22-F24). The final adaptation in the analysis plane is obtained with a vertical slit (FV) and the quadrupole doublet of the prototype (FODO structure).

Figure 7 shows the simulated vertical beam distributions at the entrance of the slit and of Q1. The slit is set to a ±10 mm opening corresponding to a vertical size of ±33 mm (FWHM) at Q1 and resembling a uniform distribution. The error concerning the filling is a few percent over the range of variation between 20 and 80%. This error in the filling evaluation will affect the accuracy of the measurements and therefore requires further investigation to enable an evaluation of the emittance growth in full detail.

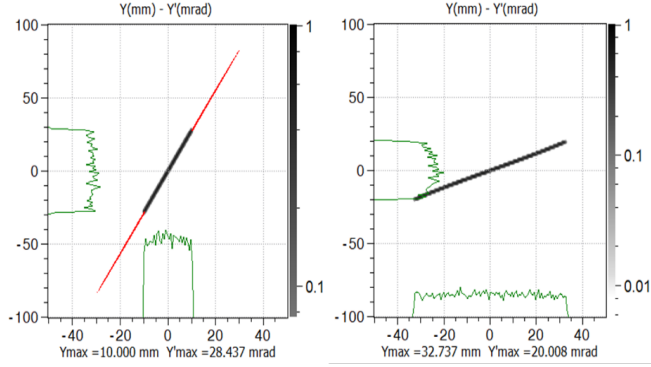

**FIG. 7.** Vertical distributions at the entrance of the FV slit (±10 mm aperture) and of the Q1 quadrupole (810 mm downstream) according to the simulations. Red points indicate the particles that have been intercepted by the slit.

The beam profile was reconstructed from the measurements of the beam currents of the FV slit and the FC Faraday cup installed at the end of the setup, see Fig. 8. Normalization of the measurements was necessary due to the difference in material/finish of the slit jaws and FC, and due to the losses in the horizontal plane associated with the smaller diameter of the FC compared to the horizontal dimension of the beam. A scale factor of 1.82 was applied to the FC reading to keep the sum of the two currents constant for small slit openings. Indeed, at small apertures, the transmission of the beam passing through the prototype is maximum and the matching between the two curves is very good, whereas at larger apertures, a fraction of the beam collides with the poles resulting in the observed losses in the horizontal plane, as indicated in Fig. 6. This confirms the importance of beam filling associated with large slit aperture and resulting beam distribution, and their potential influence on emittance growth measurement.

.

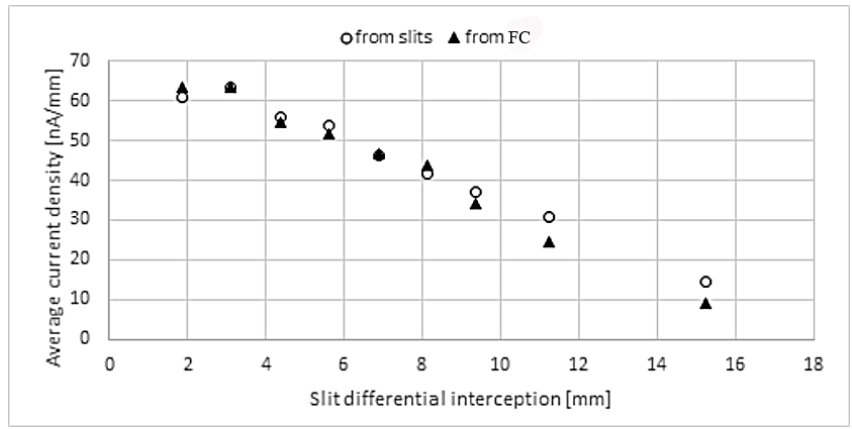

**FIG. 8.** Profile reconstruction from vertical slit (FV) and Faraday cup (FC) measurements. The discrepancy between the two series is due to beam losses occurring after the slit.

## 3.3 Emittance growth measurements in the presence of perturbations

To measure the emittance growth and the effects of the filling of the quadrupole aperture, two emittance meters (EM1 and EM2) were installed downstream of the quadrupole structure of the prototype. EM1 is a pepperpot system of Pantechnik and EM2 is an Allison type scanner [38]. Measurements were set in the vertical plane to exclude unwanted high-order contributions from the dispersion of the upstream dipoles D1 and D4 in the ARIBE line. Simulations showed that, for a single quadrupole doublet, the higher order contribution and some emittance growth should be observable at 80% filling for a beam with uniform distribution. In order to have a good control of the filling and to form an almost uniform beam distribution, the beam was strongly diverged at the vertical slit (FV) installed at 810 mm upstream of the Q1 entrance. The beam divergence and the desired settings of the quadrupoles of the ARIBE line were defined using simulations. The obtained results from the EM2 measurements are shown in Fig. 9. As expected, the beam emittance increases with slit opening, and optical aberrations increase with filling of the quadrupole but also in the presence of misalignment between the ECR ion source and the prototype. A misalignment induces an offset in position and angle in the transverse phase plane, and a shift of the emittance figure as shown in Fig. 9. It can be produced by faulty steering, internal electrical disconnections, voltage drop (large beam partially colliding with the poles), and charging of floating surfaces inside the vacuum chamber. Another scenario that needs to be studied in more detail is the possibility of a partial discharge between several poles due to the large beam filling. Therefore, deleterious effects related to parasitic steering and weak centering of the beam in the doublet structure must be avoided to allow accurate emittance growth measurements. In addition, large acceptance conditions, i.e. filling larger than 80%, can induce unstable behavior of the beam optical element.

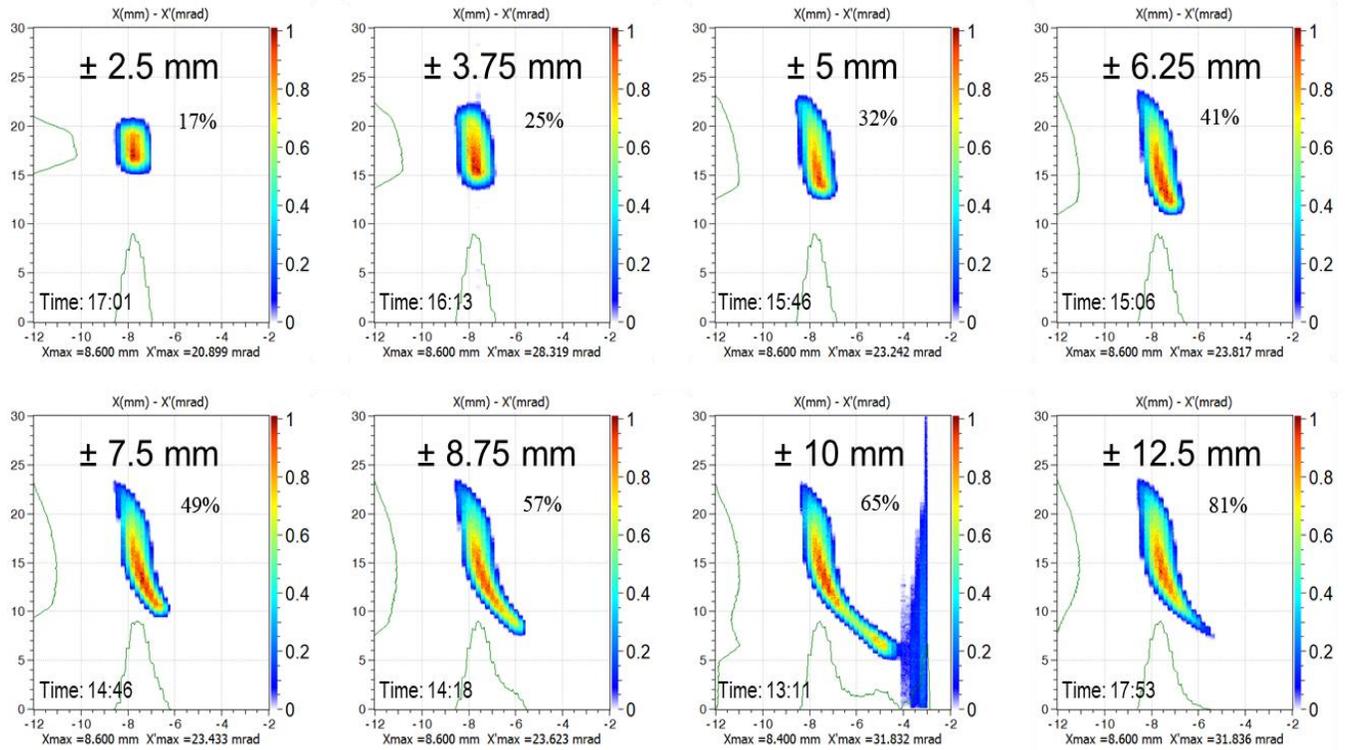

**FIG. 9.** Emittance measurements in 2D with different slit apertures and corresponding Q1 quadrupole filling (± half opening of FV slit and percentage of filling).

## 4. Discussion

As noticed during the experiment, optical elements and beam imperfections generate errors and require correction of beam orbit steering and envelope focusing. Assessing beam dynamics by accounting for errors and estimating tolerances is essential for efficient beam transport. Standard error scenarii taken for HI RIBs are 0.1 mm in position ($\Delta x,y$), and 0.1 mrad in angle ($\Delta x',y'$), see Table 1. Both static and dynamic errors affect transport performance due to their influence on beam position and size. Static errors are measurable effects corrected by steerers (orbit correction) or focusing elements (first order envelope correction). These elements allow correction of beam position and angle, misalignment and/or rotation of electromagnetic elements. On the other hand, dynamic errors are hardly measurable and are corrected with difficulty. They correspond to variation in power supply output, vibration of magnets, etc. They can be considered low compared to static errors and fixed to less than 1% of the static errors. TraceWin code simulates many types of error, defining a range for each variable, separately for static and dynamic errors. To note, to define the final aperture, the accidental errors must be considered and are occasional errors due to a rapid and final change of some variables (power shutdown, malfunction of a magnet, etc.). Without correction in position and size, deviations can exceed 100% of the initial values and correction of errors is mandatory to stay within the tolerances.

**Table 1.** Standard error input parameters and amplitudes for HI RIB transport (x,y transverse dimensions, z axial dimension).

| Beam error | |
|---|---|
| x,y position (mm) | +/- 0.1 |
| x',y' angle (mrad) | +/- 0.01 |
| x,y transverse emittance (%) | + 5 |
| dE/E energy variation (%) | +/- 0.1 |
| Optics error (focusing, deflecting, steering) | |
| x,y position (mm) | +/- 0.1 |
| x,y rotation (mrad) | +/- 0.05 |
| z rotation (mrad) | +/- 0.1 |
| dG/G gradient (%) | +/- 1 |
| Instrumentation error (Faraday cup, beam position, etc.) | |
| x,y position (mm) | +/- 0.1 |
| x,y,z rotation (mrad) | +/- 0.01 |
| Miscellaneous parameters (intensity, etc.) (%) | +/-1 |

One of the main challenges in evaluating the performance of the quadrupole doublet was related to the measurement of the signature of the HOM field components in the tail and the halo of the beam and to detect the emittance growth due to the transport through the optical element. As reported earlier, beam loss monitoring plays an important role for demanding facilities and different techniques are used to detect them, as X-ray detection, thermal probe, halo measurement, etc. See the example in [39] to control losses and check that they remain below a certain threshold by comparison between measurements (beam current, position, temperature distribution along the beam line) and simulation. With scanners that must assess emittance growth, the critical point is the spatial and angular resolution and precision. During the experiment, the resolution of the beam current intensity measurements was limited to 1pA for both units, see Fig. 12 (EM1 measurement). This barrier should clearly be pushed back in the future to allow the evaluation of emittance growth with better precision, as in [40] where the measurement of isobaric contaminants was possible with intensities reaching 2 fA.

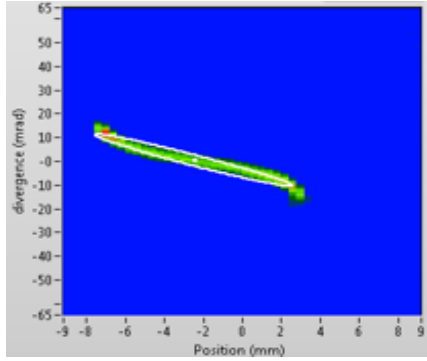

**FIG. 12.** Distribution of trace-space in the Y-Y' vertical plane obtained with the pepperpot scanner (EM1), slit opening of ±12.5 mm (FV), and 81% filling in the first quad (Q1 of QD quadrupolar doublet). Most of the beam tail and of the halo disappears due to high threshold settings and background noise filtering.

The halo is characterized by a low charge density around the core and the tail of the beam. Some typical numbers defining the halo are the particle count that is less than 10% of the total, distributed over more than 60% of the area of the 2D beam emittance, with less than $10^{-5}$ of the total beam current intensity at 5-rms for a Gaussian distribution, see references [7, 40-42]. During the experiment, the current intensity of the single charged argon beam was typically 100 nA. The beamlet resulting of the scanner sampling has a current intensity of 1 nA and the standard resolution of the electronics is 10 pA. This leads to a ratio of minimum intensity over total intensity of $10^{-4}$ reflecting the problem of measuring the emittance at 4 or 5-rms because $10^{-5}$ is required. When measuring such low intensities, the signal to noise ratio (S/N) becomes crucial and therefore the identification of the background noise (BGN), the filtering of the signal, and the reconstruction of the emittance figure keeping most of the halo will require further effort to improve emittance growth measurements. A specific algorithm was developed for the experiment to deal with non-Gaussian beam distributions and noisy measurements. Based on 2D-Gaussian and polynomial fits, it has significantly improved the accuracy of emittance measurements performed with the IPHC scanner [37]. Alternative developments include the definition of the background based on the statistical fluctuations far from the expected signal, which allows to decompose the entire emittance figure using non-analytical functions (spline fits) without constraints on the area where potential tails lie. It is foreseen to make use of artificial intelligence algorithms for image recognition of characteristic emittance shapes in the initial step and during the definition of tails and halos.

As noted earlier in reference [43], there are some inherent difficulties in measuring emittance in low-energy beam transport lines which are related to contaminants, space charge, optical aberrations, non-Gaussian distributions, and non-elliptical emittance figures which induce errors in the representation of the trace-space and in the rms calculation. Accurate measurement of absolute emittance is a difficult task and comparison between two different systems is even more difficult. A 10-20% discrepancy between two different systems is common. The experiment showed that the emittance growth must be measured at 3-rms at least with the same device, with successive focused and defocused beam, even if a comparison with a second system placed under the same conditions, located nearby and during the same run is strongly recommended. Data analysis have been carried out with a new algorithm developed specifically for this experiment allowing background noise identification, filtering, emittance figure reconstruction, and leading therefore to a better accuracy, see EM2 emittances shown in Fig. 13 and [37] for further details.

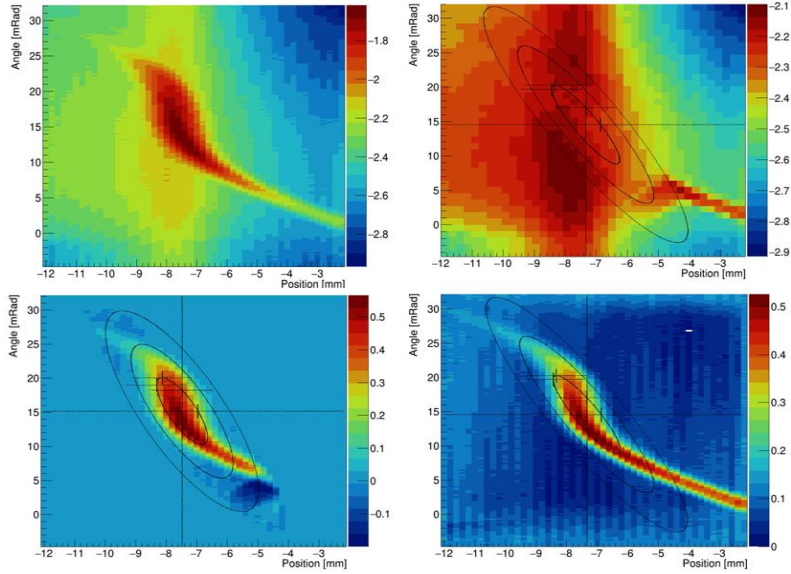

**FIG. 13.** From top left to bottom right: (a) raw data, (b) background noise identification (BGN), (c) BGN filtering, and (d) emittance figure reconstruction highlighting tails produced by optical aberrations. The measurements are performed with the IPHC scanner (Allison emittance-meter) and with the same beam line settings as in Fig. 12 (81% filling in Q1).

Other solutions are to be developed in order to allow the characterization of the prototype performance: the stability of the PS must be improved. Although current PS specifications indicate a high stability and $10^{-5}$ load regulation, behavior under load variation, with current injection into the electric circuit, induced voltage transients, and the use of a resistive load (bleeder resistor) must be investigated in more detail. This will allow control of the pole potential even with beam losses and inter-pole effects.

## 5. Conclusion

A low-energy ion beam transport prototype including a quadrupole doublet has been realized after optimization of the quadrupole design and analysis of electrical field harmonics. From simulations of the beam dynamics carried out on a unitary optical cell composed of a quadrupole doublet as well as on a virtual line composed of six identical doublets, it was observed that the minimization of some of the HOM components results in an attenuation of the optical aberrations and therefore to a reduction of the beam losses. The design can be further improved by new optimization routines developed since the first investigations. Moreover, additive manufacturing could open up new perspectives to reduce costs and enable complex pole shapes for quadrupoles. The experiment carried out with the doublet structure on an low energy ion beam line has revealed the limits of the measurements of the trace-space distributions currently possible and necessary to evaluate the tiny emittance growth generated by a unit optical cell. The importance of the signal analysis and image reconstruction correlated with background noise, non-Gaussian distributions, and non-elliptical emittance figures was demonstrated because classical statistical tools were not entirely satisfactory. Problems related to beam settings were highlighted (stability, control of the dimensions, reproducibility, uniform transverse distribution, quadrupole filling, etc.) as well as the importance of the reliability of the electric power supply (influence of the beam on the poles, beam losses, electrical discharges, and induced transients). Then, beam halo, tolerance to errors, and perturbations of orbit (due to misalignment, fluctuation, voltage drop, etc.) have to be considered in addition to static and ideal beam filling. The higher the power and the intensity, the larger the aperture of the optical element and the safety margin. With powerful beams and high RIB intensities, margin on ratio of aperture versus rms envelope can exceed 10. Depending the application and associated risk, the quadrupole aperture should be designed within a range of 1.25-rms (with a 3-rms beam filling of a Gaussian distribution) to 15-rms beam size to reduce beam losses and radio activation. The simulations require at least $10^6$ particles, and the resolution of the emittance measurements required to characterize the halo is $10^{-5}$ of the total current intensity at 5-rms for a Gaussian beam.

**Acknowledgements**

We acknowledge the experiment P1246-A granted by the GANIL iPAC and the ARIBE facility that are jointly run by the GANIL and CIMAP laboratories. We also acknowledge the 300-kV implanter of the ACACIA facility at iCUBE laboratory, CNRS/INP Strasbourg, which was used for commissioning and testing some equipment. The data analysis was partly supported by grants from the French National Agency for Research, ARRONAX-Plus n°ANR-11-EQPX-0004, IRON n°ANR-11-LABX-18-01, and ISITE NExt no ANR-16-IDE-0007. It is supported by a PhD scholarship from the Institute of Nuclear and Particle Physics (IN2P3) from the National Scientific Research Center (CNRS). F.O. wishes to thank I. Ivanenko, JINR, and W. Beeckman, Sigmaphi, for the delivery of accurate 3D field maps, and the many fruitful discussions.